\documentclass{article}

\usepackage{arxiv}

\usepackage[utf8]{inputenc} 
\usepackage[T1]{fontenc}    
\usepackage{hyperref}       
\usepackage{url}            
\usepackage{booktabs}       
\usepackage{amsfonts}       
\usepackage{nicefrac}       
\usepackage{microtype}      
\usepackage{lipsum}
\usepackage{algorithm}
\usepackage{algorithmic}
\usepackage{amsmath}

\newtheorem{Proposition}{Proposition}
\usepackage{url}
\usepackage{siunitx}
\usepackage{graphicx}
\usepackage{latexsym}
\def\qed{\hfill $\Box$}

\title{Solving Large Break Minimization Problems in a Mirrored Double Round-robin Tournament Using Quantum Annealing}

\author{
  Michiya Kuramata \\
  Department of Industrial Engineering and Economics\\
Tokyo Institute of Technology\\
Tokyo, Japan\\
\texttt{kuramata.m.aa@m.titech.ac.jp}\\
   \And
 Ryota Katsuki \\
 NTT DATA corporation\\
  Tokyo, Japan \\
 \texttt{Ryota.Katsuki@nttdata.com}\\
  \AND
  Kazuhide Nakata \\
  Department of Industrial Engineering and Economics\\
Tokyo Institute of Technology\\
Tokyo, Japan\\
\texttt{nakata.k.ac@m.titech.ac.jp}\\
}

\begin{document}
\maketitle

\section*{Abstract}
Quantum annealing has gained considerable attention because it can be applied to combinatorial optimization problems, which have numerous applications in logistics, scheduling, and finance. In recent years, with the technical development of quantum annealers, research on solving practical combinatorial optimization problems using them has accelerated. However, researchers struggle to find practical combinatorial optimization problems, for which quantum annealers outperform other mathematical optimization solvers. Moreover, there are only a few studies that compare the performance of quantum annealers with one of the most sophisticated mathematical optimization solvers, such as Gurobi and CPLEX. In our study, we determine that quantum annealing demonstrates better performance than the solvers in the break minimization problem in a mirrored double round-robin tournament. We also explain the desirable performance of quantum annealing for the sparse interaction between variables and a problem without constraints. In this process, we demonstrate that the break minimization problem in a mirrored double round-robin tournament can be expressed as a 4-regular graph. Through computational experiments, we solve this problem using our quantum annealing approach and two-integer programming approaches, which were performed using the latest quantum annealer D-Wave Advantage, and the sophisticated mathematical optimization solver, Gurobi, respectively. Further, we compare the quality of the solutions and the computational time. Quantum annealing was able to determine the exact solution in 0.05 seconds for problems with 20 teams, which is a practical size. In the case of 36 teams, it took 84.8 s for the integer programming method to reach the objective function value, which was obtained by the quantum annealer in 0.05 s. These results not only present the break minimization problem in a mirrored double round-robin tournament as an example of applying quantum annealing to practical optimization problems, but also contribute to find problems that can be effectively solved by quantum annealing.


\section{Introduction}
Quantum annealing \cite{kadowaki1998quantum} can be applied to combinatorial optimization problems, which have numerous applications in logistics, scheduling, and finance. Quantum annealers, which are hardware devices that perform quantum annealing, are relatively noise resistant and consist of an increasing number of quantum bits (qubits). In 2017, D-Wave Systems released a quantum annealer with 2048 qubits, and have now released one with 5760 qubits \cite{boothby2020next}. Although several problems \cite{dwave2021usermanual} remain unresolved, the increase in the qubits and reduction in the noise of the quantum annealer are among the primary advancements \cite{dwave2019noise}. In recent years, quantum annealers have been used to solve practical combinatorial optimization problems \cite{neukart2017traffic,nishimura2019item,inoue2021traffic,negre2020detecting,ohzeki2019control,stollenwerk2019quantum}. However, researchers struggle to find practical combinatorial optimization problems, for which quantum annealers outperform other mathematical optimization solvers \cite{ohzeki2019control,o2018nonnegative}. Moreover, only a few studies have compared the performance of quantum annealers with mathematical optimization solvers, such as Gurobi \cite{gurobi} and CPLEX \cite{cplex2009v12}. One of the fields that quantum annealing can be applied is sports scheduling. We thus considered this field to perform our comparison between quantum annealing and other solvers.

Sports scheduling involves the construction of a suitable schedule for sports competitions. It has several practical constraints and splits into various combinatorial optimization problems \cite{ribeiro2012sports,rasmussen2008round}. Among them, in a round-robin tournament (RRT), where each team plays against every other team once, in a double round-robin tournament (DRRT), where each team plays against the others twice, and in a mirrored double round-robin tournament (MDRRT), which is a DRRT with the same combination of games in the first and second halves, much research has been conducted on the break minimization problem, which determines whether the game is held in the venue of the team, or its opponent \cite{rasmussen2008round,trick2000schedule,regin2001minimization,urdaneta2018alternative,elf2003minimizing,miyashiro2006semidefinite,suzuka2007home}. RRTs, DRRTs, and MDRRTs are adopted in many professional sports, such as soccer and basketball \cite{rasmussen2008round,nemhauser1998scheduling,schreuder1992combinatorial,rasmussen2008scheduling,ribeiro2006scheduling}. The break minimization problem is derived from practical requirements, and is known to be very difficult to solve. To address this problem, methods such as integer programming \cite{trick2000schedule,urdaneta2018alternative} and constraint programming \cite{regin2001minimization} have been developed in the past.

The contributions of this study are as follows: First, we found that the break minimization problem in an MDRRT is a practical combinatorial optimization problem, which is difficult to solve using the mathematical optimization solvers and is easy to solve using quantum annealers. Second, we explain that the break minimization problem is easy to solve using quantum annealers because of the sparse interaction between variables and the lack of constraints. In this process, we demonstrate that the break minimization problem in an MDRRT can be expressed as a 4-regular graph. Third, we solve this problem using the latest quantum annealer D-Wave Advantage, and one of the most sophisticated mathematical optimization solvers, Gurobi, and compare the quality of their solutions through computational experiments. We also measure the time it takes for Gurobi to reach the objective function value, which the quantum annealer reaches in 0.05 \si{\second}.

\section{Quantum Annealing using D-Wave Advantage}
\label{sec:Quantum Annealing using D-Wave2000Q}

Quantum annealing \cite{kadowaki1998quantum} is a method for solving combinatorial optimization problems using quantum fluctuations. The solution of the combinatorial optimization problem is obtained by first applying a strong transverse field, and then gradually weakening the transverse field. This is similar to simulated annealing \cite{kirkpatrick1983optimization}; however, quantum annealing is performed using a quantum annealer as a physical phenomenon, whereas simulated annealing is calculated using a classical computer. Quantum annealing minimizes the energy of the Ising model (Eq(\ref{eq:ising model})), associated with the combinatorial optimization problem.

\begin{equation}
\label{eq:ising model}
E=\sum_{i,j} J_{i j} s_{i} s_{j}+\sum_{i} h_{i} s_{i}, \quad s_i=\pm 1,\;\; \forall i \in V.
\end{equation}

In Eq (\ref{eq:ising model}), $J_{i,j}$ is the coupling strength between the $i$th and $j$th spin, and $h_i$ is the bias. $V$ is a set of spins, and each spin $s_i$ takes the value $1$ or $-1$. By defining $s_i = 2 x_i - 1$ in Eq (\ref{eq:ising model}), we obtain quadratic unconstrained binary optimization (QUBO), which is suitable for representing combinatorial optimization problems. QUBO is defined as Eq (\ref{eq:QUBO}).

\begin{equation}
\label{eq:QUBO}
    \begin{array}{ll} \text{minimize} & \mathbf{q}^{\top} Q \mathbf{q} \\ \text {subject to} & \mathbf{q} \in \lbrace0,1\rbrace^{L}. \end{array}
\end{equation}

In Eq (\ref{eq:QUBO}), $q_i$ is a binary variable, and $L$ is the number of binary variables. $Q \in \mathbb{R}^{L \times L}$ is a matrix that characterizes the combinatorial optimization problem. QUBO is a problem that determines the values of $\mathbf{q}$ that minimize $\mathbf{q}^{\top} Q \mathbf{q}$. Because this QUBO is equivalent to the Ising model, we can solve the combinatorial optimization problem on a quantum annealer.

\begin{figure}[tb]
    \centering
    \includegraphics[keepaspectratio, scale=0.5]{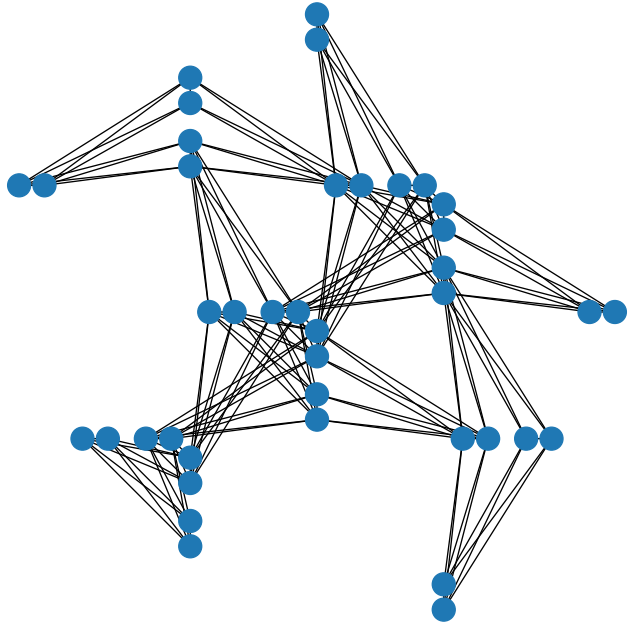}
    \caption{{\bf Pegasus graph.} The blue circles represent the qubits and the solid black lines represent the connections between the qubits.}
    \label{fig:Pegasus graph}
\end{figure}

However, to solve the Ising model transformed into QUBO using the quantum annealer, a minor-embedding is necessary. Minor embedding refers to associating the Ising model with qubits in the quantum processing unit topology (Fig \ref{fig:Pegasus graph}). Fig \ref{fig:Pegasus graph} is a Pegasus graph mounted on D-Wave Advantage, a quantum annealer with 5760 qubits. In Fig \ref{fig:Pegasus graph}, the blue circles represent the qubits and the solid black lines represent the connections between the qubits. As shown in Fig \ref{fig:Pegasus graph}, because the connectivity between qubits is sparse, multiple qubits may be required to represent one logical variable. In particular, QUBO with many non-zero elements of $Q$ in Eq (\ref{eq:QUBO}) consumes many qubits to embed it in a quantum annealer. This is difficult in practice; however, D-Wave Systems have resolved the problem. The maximum degree of the Chimera graph, the hardware graph made by D-Wave Systems, is 6. The maximum degree of the Pegasus graph, the more recent hardware graph, is 15 \cite{boothby2020next}. Therefore, it is easier to represent combinatorial optimization problems in a Pegasus graph, than in a Chimera graph.

\section{Break Minimization Problem on a Mirrored Double Round-robin Tournament}
\label{sec:Break Minimization Problem on Mirrored Double Round-robin Tournament}

\subsection{Definition of the problem}
\label{subsec:Definition of the problem}

An RRT is a competition that meets the following conditions.

\begin{enumerate}
    \item Each team meets every other team once.
    \item Each team has its own venue in its home town. A home game for a team is an away game for its opponent.
    \item Each game is played at the home of either of the teams, or its opponent.
\end{enumerate}

A DRRT is a competition that meets the following four conditions.

\begin{enumerate}
    \item Each team meets every other team twice.
    \item Each team has its own venue in its home town.
    \item Each game is played at the home of either the team or its opponent.
    \item If the first game against an opponent is played at the team's home (/ away) venue, then the second game is played at the away (/ home) venue.
\end{enumerate}
An MDRRT is a DRRT where the first half is the same as the second half, except for exchanging the home games and away games.

\begin{table}[tb]
\begin{tabular}{cc}
  \begin{minipage}[t]{.45 \textwidth}
  \centering
    \caption{\bf Timetable}
      \begin{tabular}{|l|r|r|r|r|r|r|}
      \hline
        {\bf Slot} & \bf 1 & \bf 2 & \bf 3 & \bf 4 & \bf 5 & \bf 6 \\
        \hline
        \bf team 1 &  2 &  3 &  4 &  2 &  3 &  4 \\
        \hline
        \bf team 2 &  1 &  4 &  3 &  1 &  4 &  3 \\
        \hline
        \bf team 3 &  4 &  1 &  2 &  4 &  1 &  2 \\
        \hline
        \bf team 4 &  3 &  2 &  1 &  3 &  2 &  1 \\
        \hline
      \end{tabular}
    \label{tab:timetable}
  \end{minipage}
  \hfill
  \begin{minipage}[t]{.45\textwidth}
  \centering
    \caption{\bf Home away assignment}
      \begin{tabular}{|l|r|r|r|r|r|r|}
      \hline
        {\bf Slot} &  \bf1 & \bf 2 & \bf 3 & \bf 4 & \bf 5 & \bf 6 \\
        \hline
        \bf team 1 &  1 &  0 &  1 &  0 &  1 &  0 \\
        \hline
        \bf team 2 &  0 &  0 &  1 &  1 &  1 &  0 \\
        \hline
        \bf team 3 &  1 &  1 &  0 &  0 &  0 &  1 \\
        \hline
        \bf team 4 &  0 &  1 &  0 &  1 &  0 &  1 \\
        \hline
      \end{tabular}
    \label{tab:ha_assignment}
  \end{minipage}
\end{tabular}
\end{table}

We define the symbols. The timetable shows the opponents of all teams and their slots. The slots refer to dates. A home-away assignment (HA-assignment) defines which team’s home ground, the game will be hosted at. The symbols used in the mathematical expressions are as follows.

\begin{itemize}
    \item $2n$: the number of teams. $n$ is an integer of $2$ or greater.
    \item $T \in \{1,2,\ldots,2n\}$: a set of teams.
    \item $S \in \{1,2,\ldots,2(2n-1)\}$: a set of slots.
    \item $\tau(t,s) \in T \times S$: the opponent that plays against team $t$ at slot $s$.
    \item $\mathcal{T}$: a timetable. The $(t,s)$ entry of $\mathcal{T}$ is $\tau(t,s)$.
    \item $a(t,s)$: $a(t,s)=1$ if team $t$ plays against team $\tau(t,s)$ in slot $s$ at team $t$'s home, $a(t,s)=0$ otherwise.
    \item $\mathcal{A}$: this represents an HA-assignment. The $(t,s)$ entry of $\mathcal{A}$ is $a(t,s)$.
\end{itemize}

Table \ref{tab:timetable} is a timetable $\mathcal{T}$ for $n=2$. In Table \ref{tab:timetable}, the vertical axis represents the teams and the horizontal axis represents the slots. Each entry $(t,s)$ of the timetable $\mathcal{T}$ is an opponent $\tau(t,s)$. Because Table \ref{tab:timetable} is the timetable for the MDRRT, $\tau(t,s)=\tau(t,s+2n-1) \quad (\forall (t,s) \in T \times \{1,\ldots,2n-1\})$ holds. Table \ref{tab:ha_assignment} shows the HA-assignment $\mathcal{A}$ corresponding to the timetable in Table \ref{tab:timetable}. Each entry $(t,s)$ of $\mathcal{A}$ indicates whether a game in which team $t$ plays against team $\tau(t,s)$ in slot $s$ is a home game or an away game. The game between team $t$ and team $\tau(t,s)$ in slot $s$ is held at the home of team $t$ if $a(t,s)=1$ and at the home of team $\tau(t,s)$ if $a(t,s)=0$. A break means that a team plays at two home games or two away games in a row. For example, in Table \ref{tab:ha_assignment}, Team 2 has a break in slot 2 because of the two consecutive away games. Similarly, Team 2 has a break in slot 4 because of the two consecutive home games. Breaks should be avoided as much as possible to ensure fairness among teams.

As described in \cite{nemhauser1998scheduling} and  \cite{schreuder1992combinatorial}, in practical application, various constraints must be considered in scheduling sports tournaments. Further, R{\'e}gin \cite{regin2001minimization} divided the scheduling in an RRT into a first stage, where many practical constraints are involved, and a second stage, where there are no constraints; the first and second stages are as follows.

\begin{itemize}
    \item Considering various constraints, the schedule is created without deciding the assignment of the venue.
    \item Each game is assigned to either the home game or away game.
\end{itemize}

This approach is useful when there are many constraints in the creation of a schedule. We also refer to the problem of finding an HA-assignment with the smallest number of breaks for a given timetable as a break minimization problem. This corresponds to the second stage. Trick \cite{trick2000schedule} points out that solving the break minimization problem is more difficult than determining the timetable corresponding to the first stage. In this study, we focus on solving the break minimization problem, as well as R{\'e}gin \cite{regin2001minimization} and Trick \cite{trick2000schedule}.

\subsection{Previous studies}
\label{subsec:Previous studies}
The break minimization problem in an RRT has been studied by many researchers \cite{trick2000schedule,urdaneta2018alternative,miyashiro2006semidefinite}. In particular, De Werra and Dominique \cite{de1981scheduling} proved that the number of breaks in an RRT is more than $2n-2$, and the number of breaks in an MDRRT is more than $6n-6$. As explained in Section \ref{subsec:Definition of the problem}, $2n$ is the number of teams. Although these properties were clarified, it is very difficult to solve the break minimization problem in an RRT, a DRRT, and an MDRRT. To solve this problem, constraint programming  \cite{regin2001minimization}, integer programming \cite{trick2000schedule,urdaneta2018alternative}, and an approximation algorithm \cite{miyashiro2006semidefinite} have been studied in the past. In recent years, Urdaneta \textit{et al.} \cite{urdaneta2018alternative} demonstrated, through numerical experiments that formulating the problem as an unconstrained quadratic integer programming problem and solving it using the mathematical optimization solver is superior to other formulations with constraints. In this section, we introduce the study by Urdaneta \textit{et al.} \cite{urdaneta2018alternative}.

Before describing formulations in \cite{urdaneta2018alternative}, we explain certain symbols required in the formulations.
\begin{equation}
\label{eq:mathcal_K}
    \mathcal{K}(k)=\lbrace  (t_{k},s_{k}),(t_{k},s_{k}^{'}),(t_{k}^{'},s_{k}),(t_{k}^{'},s_{k}^{'}) \rbrace
\end{equation}

In Eq (\ref{eq:mathcal_K}), $\mathcal{K}(k)$ represents the games and their slots between the two teams of the $k$th combination. $k$ can be any integer between $1$ and ${}_{2n}\mathrm{C}_2$. $t_k$ and $t_k^{\prime}$ represent the two teams in the $k$th combination, respectively. The slot of the first game between $t_k$ and $t_k^{\prime}$ is $s_k$, and the slot of the second game is $s_k^{\prime}$. For any $k$, Eq (\ref{eq:threecondition}) holds.

\begin{equation}
\label{eq:threecondition}
    |\mathcal{K}(k) |=4,\quad k\neq k^{'} \Rightarrow \mathcal{K}(k) \cap \mathcal{K}(k^{'}) = \emptyset, \quad T \times S = \bigcup_{k=1}^{n(2n-1)}\mathcal{K}(k).
\end{equation}

$ |\mathcal{K}(k) |=4$ indicates that $ \mathcal{K}(k) $ contains four games in the timetable. $k\neq k^{'} \Rightarrow \mathcal{K}(k) \cap \mathcal{K}(k^{'}) = \emptyset$ also indicates that two different $\mathcal{K}(k)$ do not contain the same games. $T \times S = \bigcup_{k=1}^{n(2n-1)}\mathcal{K}(k)$ indicates that adding up all the combinations equals the original timetable.

The binary variable $y_{ts}$ takes $1$ if team $t$ plays at home in slot $s$ and 0 if it plays away. Eq (\ref{eq:constraint of DRRT}) holds for any $k \in \{1,2,\ldots,n(2n-1)\}$ to satisfy the conditions of the DRRTs.

\begin{equation}
\label{eq:constraint of DRRT}
\begin{array}{l}
y_{t_{k} s_{k}}+y_{t_{k}^{\prime} s_{k}}=1 \\
y_{t_{k} s_{k}}+y_{t_{k} s_{k}^{\prime}}=1 \\
-y_{t_{k} s_{k}}+y_{t_{k}^{\prime} s_{k}^{\prime}}=0
\end{array}
\end{equation}

The first constraint is that each team plays at its home or at its opponent's home. The second constraint implies that a team plays at home (/ away) against its opponent first, and then plays away (/ home) against its opponent second. The third implies that the first game is played at the home (/ away) of the team, and the second game is played at the home (/ away) of the opponent. These represent the second, third, and fourth of the four conditions for a DRRT, respectively.

Urdaneta \textit{et al.} did not provide a specific formulation for the break minimization problem. They demonstrated that the constrained optimization problem (Eq (\ref{eq:urdaneta1})), which represents a break minimization problem in a DRRT, can be expressed as an unconstrained optimization problem (Eq (\ref{eq:urdaneta2})).

\begin{equation}
\label{eq:urdaneta1}
\begin{array}{ll}
\text{minimize} & l(\mathbf{y})=c^{\top} \mathbf{y}+\frac{1}{2} \mathbf{y}^{\top} H \mathbf{y} \\
\text { subject to } 
& y_{t_{k} s_{k}}+y_{t_{k}^{\prime} s_{k}}=1 \quad (\forall k \in \{1,2,\ldots,n(2n-1)\}),\\
& y_{t_{k} s_{k}}+y_{t_{k} s_{k}^{\prime}}=1 \quad (\forall k \in \{1,2,\ldots,n(2n-1)\}),\\
& -y_{t_{k} s_{k}}+y_{t_{k}^{\prime} s_{k}^{\prime}}=0 \quad (\forall k \in \{1,2,\ldots,n(2n-1)\}),\\
& y_{t, s} \in\{0,1\} \quad(\forall t \in \mathrm{T}, \forall s \in \mathrm{S}).
\end{array}
\end{equation}

The constraints are the same as in Eq (\ref{eq:constraint of DRRT}). We define $z_k$ as $z_k := y_{t_k s_k}$ for the first component $(t_k,s_k)$ of $\mathcal{K}(k)$. The variables relating to the components of $\mathcal{K}(k)$ can be replaced, as in Eq (\ref{eq:variable transformation}).

\begin{equation}
\label{eq:variable transformation}
    y_{t_k s_k} = z_k,\quad y_{t_k^{\prime} s_k} = 1 - z_k,\quad  y_{t_k s_k^{\prime}} = 1 - z_{k},\quad y_{t_k^{\prime} s_k^{\prime}} = z_k
\end{equation}

Thus, substituting Eq (\ref{eq:variable transformation}) into Eq (\ref{eq:urdaneta1}) yields Eq (\ref{eq:urdaneta2}).

\begin{equation}
\label{eq:urdaneta2}
\begin{array}{ll}
\text { minimize } & \bar{l}(\mathbf{z})=\bar{a}+\bar{c}^{\top} \mathbf{z}+\frac{1}{2} \mathbf{z}^{\top} \bar{H} \mathbf{z} \\
\text { subject to } & \mathbf{z} \in\{0,1\}^{n(2 n-1)}.
\end{array}
\end{equation}

In Eq (\ref{eq:urdaneta2}), $\bar{a}$, $\bar{c}$, and $\bar{H}$ are obtained by substituting Eq (\ref{eq:variable transformation}) for Eq (\ref{eq:urdaneta1}).

\section{Formulation and Analysis of the Problem}
In Section \ref{subsec:Formulation Break Minimization Problem on Mirrored Double Round-robin Tournament}, we specify the formulation by Urdaneta \textit{et al.} \cite{urdaneta2018alternative} as a break minimization problem in an MDRRT. In Section \ref{sec:Analysis:The Benefits of the sparsity of the problem} and \ref{sec:Analysis:The Benefits of No Constraints}, we explain that a break minimization problem in an MDRRT is easy to solve using a quantum annealer.

\subsection{Formulating the break minimization problem in a mirrored double round-robin tournament}
\label{subsec:Formulation Break Minimization Problem on Mirrored Double Round-robin Tournament}
As mentioned earlier, Urdaneta \textit{et al.} \cite{urdaneta2018alternative} does not provide a specific objective function for Eq (\ref{eq:urdaneta1}) and Eq (\ref{eq:urdaneta2}). Further, we set the objective function as the total number of breaks in the MDRRT. Our formulation corresponding to Eq (\ref{eq:urdaneta1}) is Eq (\ref{eq:myformulation1}).

\begin{equation}
\label{eq:myformulation1}
\begin{array}{ll}
\text{minimize} & f(\mathbf{y})=\sum_{t \in T}\sum_{s \in S \setminus \{4n-2\}}(y_{t s}y_{t s+1}+(1-y_{t s})(1-y_{t s+1})) \\
\text{ subject to }
& y_{t_{k} s_{k}}+y_{t_{k}^{\prime} s_{k}}=1 \quad (\forall k \in \{1,2,\ldots,n(2n-1)\}),\\
& y_{t_{k} s_{k}}+y_{t_{k} s_{k}^{\prime}}=1 \quad (\forall k \in \{1,2,\ldots,n(2n-1)\}),\\
& -y_{t_{k} s_{k}}+y_{t_{k}^{\prime} s_{k}^{\prime}}=0 \quad (\forall k \in \{1,2,\ldots,n(2n-1)\}),\\
& y_{t s} \in\{0,1\} \quad(\forall t \in \mathrm{T}, \forall s \in \mathrm{S}).
\end{array}
\end{equation}

The objective function $f(\mathbf{y})$ consists of the sum of $y_{t s}y_{t,s+1}$, which represents a break corresponding to the two consecutive home games and $(1-y_{t s})(1-y_{t s+1})$, which represents a break corresponding to the two consecutive away games. As Urdaneta \textit{et al.} did, we use transform Eq (\ref{eq:myformulation1}) into Eq (\ref{eq:myformulation2}) using Eq (\ref{eq:variable transformation}).

\begin{equation}
\label{eq:myformulation2}
\begin{array}{ll}
\text{minimize} & \sum_{t_k \in T}\sum_{s_k \in S \setminus \{4n-2\}} a(t_k,s_k) \left(z_k z_{k^{\prime}}+(1-z_k)(1-z_{k^{\prime}})\right)\\ & \hspace{23mm}+ b(t_k,s_k) \left((1 - z_k) z_{k^{\prime}}+z_k (1-z_{k^{\prime}})\right)\\ &\hspace{23mm}+ c(t_k,s_k)\left(z_k (1 - z_{k^{\prime}})+(1-z_k)z_{k^{\prime}}\right) \\&\hspace{23mm}+  d(t_k,s_k)\left(z_k z_{k^{\prime}}+ (1 - z_k)(1 - z_{k^{\prime}})\right)\\
\text { subject to }
& z_k \in \{0,1\} \quad (\forall k \in \{1,2,\ldots,n(2n-1)\}).
\end{array}
\end{equation}

In Eq (\ref{eq:myformulation2}), $k^{\prime}$ satisfies $(t_k,s_k + 1) \in \mathcal{K}(k^{\prime})$ for any $k$. Let $a(t_k,s_k)=1$ if $(y_{t_k s_k}=z_k) \land  (y_{t_k s_k + 1}=z_{k^{\prime}})$; otherwise, $a(t_k,s_k)=0$. Let $b(t_k,s_k)=1$ if $(y_{t_k s_k}= 1 - z_k) \land (y_{t_k s_k + 1}=z_{k^{\prime}})$, otherwise $b(t_k,s_k)=0$. Let $c(t_k,s_k)=1$ if $(y_{t_k s_k}=z_k)\land (y_{t_k s_k + 1}= 1 - z_{k^{\prime}})$, otherwise $c(t_k,s_k)=0$. Let $d(t_k,s_k)=1$ if $(y_{t_k s_k}= 1 - z_k) \land (y_{t_k s_k + 1}= 1 - z_{k^{\prime}})$, otherwise $d(t_k,s_k)=0$. Eq (\ref{eq:myformulation2}) is the QUBO formulation. Therefore, we transformed Eq (\ref{eq:myformulation2}) into an equivalent Ising model and solved it using quantum annealing. As described in Section \ref{sec:Analysis:The Benefits of the sparsity of the problem}, the break minimization problem in an MDRRT is considerably suitable for solving the problem using a quantum annealer.

\subsection{Analysis : Benefits of the sparsity of the problem}
\label{sec:Analysis:The Benefits of the sparsity of the problem}

\begin{figure}[tb]
    \centering
    \includegraphics[keepaspectratio, scale=0.5]{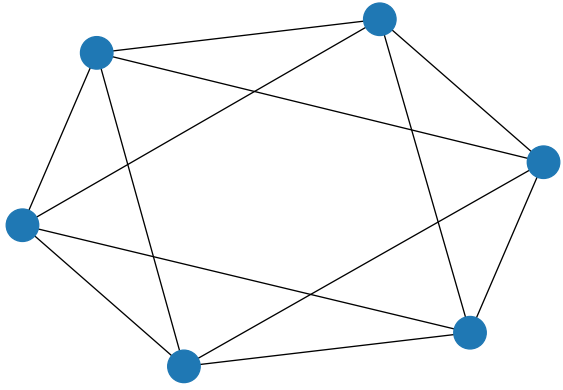}
    \caption{ {\bf Source graph of the break minimization problem defined by Table \ref{tab:timetable}.}
    The blue nodes represent logical variables, and the solid black lines represent the interaction between variables.}
    \label{fig:source_graph}
\end{figure}

In this section, we explain that a break minimization problem in an MDRRT is easy to solve using a quantum annealer, because the graph representation of the problem is sparse. As in  \cite{dwave2021ocean}, we refer to the graph representation of QUBO as the source graph, and the quantum processing unit topology of a quantum annealer as the target graph. The source graph contains nodes that represent logical variables and edges that represent the interaction between variables. The target graph has nodes that mean qubits and edges, which indicate the connections between the qubits, as shown in Fig \ref{fig:Pegasus graph}. Fig \ref{fig:source_graph} shows the source graph of the break minimization problem for the schedule defined in Table \ref{tab:timetable} and Table \ref{tab:ha_assignment}. As can be observed from Fig \ref{fig:source_graph}, the source graph of the break minimization problem in an MDRRT is a 4-regular graph.

\begin{Proposition}
The source graph of the break minimization problem in an MDRRT is a 4-regular graph.
\end{Proposition}
\textit{Proof} As shown in Eq (\ref{eq:variable transformation}),  the variable $z_k$ represents the four games denoted by $\{y_{ts} |(t,s)\in \mathcal{K}(k)\}$. In addition, in an MDRRT, the combination of opponents for the games before and after a certain game is the same for the first and second halves, and thus, $z_k$ is related to the four variables. Therefore, the source graph of a break minimization problem in an MDRRT is a 4-regular graph. \qed

Regardless of the number of teams in an MDRRT, the source graph is a 4-regular graph, which means that the graph is very sparse. However, the source graph in a DRRT is not always a regular graph, and its degree is less than, or equal to 8.

\begin{Proposition}
The degree of the source graph of a break minimization problem in a DRRT is less than or equal to 8.
\end{Proposition}
\textit{Proof}
The variable $z_k$ represents the four games denoted by $\{y_{ts} |(t,s)\in \mathcal{K}(k)\}$. In contrast to an MDRRT, in a DRRT, the combination of opponents in the games before and after a certain game is not always equal in the first and second halves. Therefore, when the degree of the source graph is the largest, $z_k $ is related to eight variables that represent the before and after of the game in the first and second halves. Thus, the maximum degree of the source graph of the break minimization problem in a DRRT is 8.\qed

In Tables \ref{tab:nodes_edges} and \ref{tab:nodes_edges_drrt}, we demonstrate the minor embeddings of the source graph of the break minimization problem in an MDRRT and a DRRT, respectively. The minor embeddings are in the target graph of the D-Wave Advantage. We perform minor embedding using the method described in [1]. We randomly create five break minimization problems in both, MDRRTs and DRRTs. Tables \ref{tab:nodes_edges} and \ref{tab:nodes_edges_drrt} show the number of nodes (\textbf{Nodes} in the tables) and edges (\textbf{Edges} in the tables) in the source graph of the break minimization problem in an MDRRT and a DRRT, the total number of qubits used (\textbf{Qubits} in the tables), and the number of qubits used per node in the source graph (\textbf{Qubits/Nodes} in the tables). \textbf{Teams} in the tables represent the number of teams included in the problems.
In the target graph of D-Wave Advantage, we were able to minor-embed up to 48 teams in an MDRRT, and up to 28 teams in a DRRT. Therefore, up to 48 teams can be considered in Table \ref{tab:nodes_edges}, and up to 28 teams can be considered in Table \ref{tab:nodes_edges_drrt}. In Tables \ref{tab:nodes_edges} and \ref{tab:nodes_edges_drrt}, we can observe that the break minimization problem in an MDRRT and a DRRT is sparse. Comparing Table \ref{tab:nodes_edges} and Table \ref{tab:nodes_edges_drrt}, we can observe that the source graph in an MDRRT is sparser than that in a DRRT, and the number of qubits used is smaller. In  \cite{hamerly2019experimental}, the authors investigated how the degree of the graph of MAX-CUT problems affects the quality of the solution obtained from a quantum annealer, and demonstrated that the smaller the degree of the graph, the better the quality. Thus, it is expected that the break minimization problem in an MDRRT is easier to solve, than in a DRRT, using a quantum annealer. This is also confirmed by the experimental results shown in Tables \ref{tab:experiment2} and \ref{tab:experiment2_drrt}. The size of the problem that can be solved on the quantum annealer is also larger for an MDRRT than for a DRRT. We were able to solve problems with up to $48$ teams in an MDRRT and up to $28$ teams in a DRRT, using D-Wave Advantage.

\begin{table}[tb]
  \begin{minipage}[t]{.45\textwidth}
   \caption{\bf Number of nodes and edges of the source graph and used qubits in MDRRTs}
        \begin{tabular}{|l|r|r|r|r|}
        \hline
        {\bf Teams} &  \bf Nodes &  \bf Edges & \bf Qubits & \bf Qubits/Nodes \\
        \hline
        4  &      6 &     12 &       8 &      1.333333 \\
        \hline
        8  &     28 &     56 &      39 &      1.392857 \\
        \hline
        12 &     66 &    132 &     112 &      1.696970 \\
        \hline
        16 &    120 &    240 &     209 &      1.741667 \\
        \hline
        20 &    190 &    380 &     370 &      1.947368 \\
        \hline
        24 &    276 &    552 &     547 &      1.981884 \\
        \hline
        28 &    378 &    756 &     755 &      1.997354 \\
        \hline
        32 &    496 &    992 &    1467 &      2.957661 \\
        \hline
        36 &    630 &   1260 &    1663 &      2.639683 \\
        \hline
        40 &    780 &   1560 &    1973 &      2.529487 \\
        \hline
        44 &    946 &   1892 &    3215 &      3.398520 \\
        \hline
        48 &   1128 &   2256 &    3975 &      3.523936 \\
        \hline
        \end{tabular}
        \label{tab:nodes_edges}
  \end{minipage}
  \hfill
  \begin{minipage}[t]{.45\textwidth}
    \caption{\bf Number of nodes and edges of the source graph and qubits used in DRRTs}
        \begin{tabular}{|l|r|r|r|r|}
        \hline
        {\bf Teams} & \bf Nodes &  \bf Edges & \bf Qubits & \bf Qubits/Nodes \\
        \hline
        4  &    6 &    10.4 &     7.2 &      1.200000 \\
        \hline
        8  &   28 &    83.2 &    50.8 &      1.814286 \\
        \hline
        12 &   66 &   232.8 &   212.0 &      3.212121 \\
        \hline
        16 &  120 &   441.6 &   613.2 &      5.110000 \\
        \hline
        20 &  190 &   676.0 &  1167.8 &      6.146316 \\
        \hline
        24 &  276 &   998.4 &  2412.4 &      8.740580 \\
        \hline
        28 &  378 &  1411.2 &  3878.4 &     10.260317 \\
        \hline
        \end{tabular}
        \label{tab:nodes_edges_drrt}
  \end{minipage}
  
\end{table}

\subsection{Analysis: Benefits of no constraints}
\label{sec:Analysis:The Benefits of No Constraints}
In this section, we explain that the break minimization problem in an MDRRT is easy to solve using a quantum annealer because it is an unconstrained optimization problem.
As shown in Eq (\ref{eq:myformulation2}), the break minimization problem can be expressed naturally as an unconstrained optimization problem. In other words, all solutions searched for by quantum annealing are feasible solutions, and thus, the search conducted using quantum annealing is efficient. However, for problems with hard constraints, such as quadratic assignment problems \cite{koopmans1957assignment} and traveling salesman problems \cite{dantzig1954solution}, many of the solutions explored using quantum annealing are deemed infeasible. As an example, we consider an optimization problem with severe constraints (Eq (\ref{eq:onehot})), such as the quadratic assignment problem or traveling salesman problem, wherein the objective function is assumed to return 0 regardless of the solution.

\begin{equation}
\label{eq:onehot}
\begin{array}{ll}
\text{minimize} & 0 \\
\text { subject to }
& \sum_{i=1}^{n} x_{ij} = 1 \quad ( \forall j \in \{1,2,\ldots,n \} ), \\
& \sum_{j=1}^{n} x_{ij} = 1 \quad ( \forall i \in \{1,2,\ldots,n \} ), \\
& x_{ij} \in\{0,1\} \quad(\forall i \in \{1,2,\ldots,n\}, \forall j \in \{1,2,\ldots,n\}).
\end{array}
\end{equation}

By transforming the optimization problem (Eq (\ref{eq:onehot})) into the QUBO formulation, we obtain Eq (\ref{eq:onehot_qubo}).

\begin{equation}
\label{eq:onehot_qubo}
\begin{array}{ll}
\text{minimize} & \sum_{i=1}^{n}\left (1 - \sum_{j=1}^{n}x_{ij} \right)^2 +\sum_{j=1}^{n}\left (1 - \sum_{i=1}^{n}x_{ij} \right)^2 \\
\text { subject to }
& x_{ij} \in\{0,1\} \quad(\forall i \in \{1,2,\ldots,n\}, \forall j \in \{1,2,\ldots,n\}).
\end{array}
\end{equation}

In this case, $n^2$ binary variables are required. The set of solutions explored using quantum annealing has $2^{n^2}$ elements. However, the set of feasible solutions has $n!$ elements. Therefore, the set of feasible solutions is exponentially smaller than the set of solutions explored using quantum annealing, which makes the search inefficient. We solved Eq (\ref{eq:onehot_qubo}) using the quantum annealer D-Wave Advantage and obtain 10000 solutions. The results are summarized in Table \ref{tab:feasible}. In Table \ref{tab:feasible}, \textbf{per\_feasible} is the percentage of feasible solutions, \textbf{ev\_break} is the average number of violated constraints, and \textbf{ev\_energy} is the average value of the objective function (Eq (\ref{eq:onehot_qubo})). From Table \ref{tab:feasible}, we can observe that as the problem size $n$ increases, the number of feasible solutions decreases. This is also consistent with the result obtained by solving the quadratic assignment problem on a quantum annealer \cite{kuramata2021larger}. Thus, these results demonstrate that it is difficult to solve a hard-constrained optimization problem using a quantum annealer. However, unconstrained optimization problems do not have this feasibility problem. It is clear that unconstrained problems are easier to solve using quantum annealing.

\begin{table}[tb]
\centering
\caption{\textbf{ Percentage of feasible solutions obtained.} The parameters of the D-Wave Advantage are as follows: \textit{annealing\_time}= 50 \si{\micro}\si{\second}. \textit{num\_reads}=10000}
\label{tab:feasible}
\begin{tabular}{|l|r|r|r|}
\hline
{\bf n} & \bf per\_feasible &  \bf ev\_break & \bf ev\_energy \\
\hline
2  &      0.999500 &   0.001000 &   0.001000 \\
\hline
3  &      0.988900 &   0.022600 &   0.022600 \\
\hline
4  &      0.807642 &   0.390735 &   0.390735 \\
\hline
5  &      0.384660 &   1.460398 &   1.470503 \\
\hline
6  &      0.091352 &   2.935288 &   2.986879 \\
\hline
7  &      0.011900 &   4.791000 &   5.017400 \\
\hline
8  &      0.001800 &   6.198600 &   6.604800 \\
\hline
9  &      0.000000 &   8.187100 &   9.231800 \\
\hline
10 &      0.000000 &  11.578900 &  14.923600 \\
\hline
11 &      0.000000 &  11.746400 &  15.115000 \\
\hline
12 &      0.000000 &  11.852400 &  14.702600 \\
\hline
\end{tabular}
\end{table}

\section{ Numerical experiments and Discussion}
\label{sec:Result and Discussion}
The break minimization problem is a problem of finding an HA-assignment that minimizes the number of breaks for a given timetable. We conducted two experiments to compare our method based on quantum annealing with two integer programming approaches \cite{trick2000schedule,urdaneta2018alternative}, which exhibited excellent results. In the first experiment, we solved the break minimization problem in an MDRRT using quantum annealing, and two other integer programming approaches presented by Urdaneta \textit{et al.} \cite{urdaneta2018alternative} and Trick \cite{trick2000schedule}, respectively. Further, we compared the quality of the solutions and the computational time. In the second experiment, we measured the time it took for the integer programming approach given by Urdaneta \textit{et al.} to reach the objective function value, which the quantum annealer obtained in 0.05 \si{\second}. To demonstrate the advantage of the sparse source graph, we also conducted the same experiments for the break minimization problem in a DRRT, as well as an MDRRT. Table \ref{tab:experiment1_drrt} corresponds to Table \ref{tab:experiment1}, and Table \ref{tab:experiment2_drrt} corresponds to Table \ref{tab:experiment2}.

The timetables of the MDRRTs used in the computational experiments were created using the following method.

\begin{enumerate}
    \item Creating an RRT timetable using the Kirkman method  \cite{kirkman1847on}.
    \item Shuffling the order of the slots in the timetable created in step 1.
    \item Concatenating the two identical timetables that are shuffled in step 2.
\end{enumerate}

In this manner, we created five timetables for each number of teams and used common timetables in both computational experiments. The experimental results (Table \ref{tab:experiment1},\ref{tab:experiment1_drrt},\ref{tab:experiment2}, and \ref{tab:experiment2_drrt}) show the average of the results obtained by solving the break minimization problems defined by the five timetables. In the computational experiments, we compared the following three methods:

\begin{itemize}
    \item QA: we solve Eq (\ref{eq:myformulation2}) with quantum annealing.
    \item IP(Urdaneta): we solve Eq (\ref{eq:myformulation2}) using the integer programming approach by  Urdaneta \textit{et al.}  \cite{urdaneta2018alternative}, and we use Gurobi \cite{gurobi} as an integer quadratic programming solver.
    \item IP(Trick): we solve the problem that equals to the break minimization problem with integer programming approach by Trick  \cite{trick2000schedule}, and we use Gurobi \cite{gurobi} as an integer linear programming solver.
\end{itemize}

QA, IP(Urdaneta), and IP(Trick) are abbreviations for each method. Although all three methods were compared in the first experiment (Table \ref{tab:experiment1},\ref{tab:experiment1_drrt}), we compared our quantum annealing approach with the integer programming approach presented by Urdaneta  (IP(Urdaneta)) in the second experiment (Table \ref{tab:experiment2},\ref{tab:experiment2_drrt}) because the integer programming approach presented by Urdaneta \textit{et al.} is superior to that of Trick in the first experiment.

The computational environment for IP(Urdaneta) and IP(Trick) is as follows: we used the Gurobi Optimizer(version 9.1.2) on an Intel Core i7-7700HQ 2.80 GHz CPU with four cores and eight threads. We terminated the computation by Gurobi after 300 \si{\second} because we were required to solve a large number of problems. The parameters for QA are as follows. We used the quantum annealer, D-Wave Advantage. The version of D-Wave Advantage is Advantage\_system1.1  and has 5760 qubits. The annealing\_time is 50 \si{\micro} \si{\second} and num\_reads is 1000. Therefore, the execution time was $0.05$ \si{\second}. We minor-embed the problem using the method \cite{cai2014practical} onto the D-Wave Advantage. 

The abbreviations in Tables \ref{tab:experiment1} and \ref{tab:experiment1_drrt} are as follows: \textbf{Teams} is the number of teams, \textbf{Breaks} is the average number of breaks in the solution obtained using each method, and \textbf{Time} is the average computational time. \textbf{OPTIMAL} is the percentage of optimal solutions obtained in 300 \si{\second} using integer programming approaches. For example, \textbf{OPTIMAL} = 0.2 means that one optimal solution out of five was obtained. The abbreviations in Tables \ref{tab:experiment2} and \ref{tab:experiment2_drrt} are as follows: \textbf{Teams} is the number of teams. \textbf{Breaks(QA)} are the average number of breaks in the solution obtained from the quantum annealer. \textbf{Time(Urdaneta)} is the average computational time required for IP(Urdaneta) to reach the number of breaks in the solution obtained from the quantum annealer in 0.05 \si{\second}.

\begin{table}[ht!]
\centering
\caption{\bf QA vs. IP in MDRRTs}
\label{tab:experiment1}
\begin{tabular}{|l|r|r|r|r|r|r|r|r|}
\hline
{} & \multicolumn{2}{|l|}{\bf QA} & \multicolumn{3}{|l|}{\bf IP(Urdaneta)} & \multicolumn{3}{|l|}{\bf IP(Trick)} \\
\hline
{\bf Teams} & \bf Breaks & \bf Time(s) &  \bf Breaks &\bf Time(s) & \bf OPTIMAL & \bf Breaks &\bf  Time(s) &\bf OPTIMAL \\
\hline
4                       &        6.0 &             0.05 &          6.0 &         0.033974 &        1.0 &        6.0 &         0.039900 &        1.0 \\
\hline
8                       &       19.6 &             0.05 &         19.6 &         0.063883 &        1.0 &       19.6 &         0.288907 &        1.0 \\
\hline
12                      &       38.8 &             0.05 &         38.8 &         0.157146 &        1.0 &       38.8 &         1.951506 &        1.0 \\
\hline
16                      &       66.0 &             0.05 &         66.0 &         0.681240 &        1.0 &       66.0 &        40.549676 &        1.0 \\
\hline
20                      &      106.8 &             0.05 &        106.8 &         3.449914 &        1.0 &      106.8 &       300.039447 &        0.0 \\
\hline
24                      &      161.6 &             0.05 &        156.4 &        52.528646 &        1.0 &         -  &               -  &         -  \\
\hline
28                      &      224.8 &             0.05 &        214.0 &       252.368946 &        0.2 &         -  &               -  &         -  \\
\hline
32                      &      280.4 &             0.05 &        267.2 &       288.295851 &        0.2 &         -  &               -  &         -  \\
\hline
36                      &      368.8 &             0.05 &        346.0 &       300.026792 &        0.0 &         -  &               -  &         -  \\
\hline
40                      &      453.6 &             0.05 &        422.4 &       300.032120 &        0.0 &         -  &               -  &         -  \\
\hline
44                      &      553.6 &             0.05 &        520.8 &       300.024345 &        0.0 &         -  &               -  &         -  \\
\hline
48                      &      663.6 &             0.05 &        618.8 &       300.024338 &        0.0 &         -  &               -  &         -  \\
\hline
\end{tabular}
\end{table}

\begin{table}[tb]
\centering
\caption{\bf QA vs. IP in DRRTs}
\label{tab:experiment1_drrt}
\begin{tabular}{|l|r|r|r|r|r|r|r|r|}
\hline
{} & \multicolumn{2}{|l|}{\bf QA} & \multicolumn{3}{|l|}{\bf IP(Urdaneta)} & \multicolumn{3}{|l|}{\bf IP(Trick)} \\
\hline
{\bf Teams} & \bf Breaks & \bf Time(s) & \bf  Breaks & \bf Time(s) & \bf OPTIMAL & \bf Breaks & \bf Time(s) &  \bf OPTIMAL \\
\hline
4                       &        5.6 &             0.05 &          5.6 &         0.035595 &        1.0 &        5.6 &         0.028979 &        1.0 \\
\hline
8                       &       26.8 &             0.05 &         26.8 &         0.089388 &        1.0 &       26.8 &         0.560439 &        1.0 \\
\hline
12                      &       65.6 &             0.05 &         65.6 &         0.645643 &        1.0 &       65.6 &        10.696434 &        1.0 \\
\hline
16                      &      116.4 &             0.05 &        113.2 &        89.337958 &        0.8 &      114.4 &       300.021179 &        0.0 \\
\hline
20                      &      184.4 &             0.05 &        173.6 &       237.070479 &        0.4 &      256.0 &       300.028045 &        0.0 \\
\hline
24                      &      276.0 &             0.05 &        247.6 &       300.023959 &        0.0 &       -    &           -      &        -   \\
\hline
28                      &      409.6 &             0.05 &        362.0 &       300.023397 &        0.0 &       -    &           -      &        -  \\
\hline
\end{tabular}
\end{table}

\begin{table}[tb]
\begin{tabular}{cc}
  \begin{minipage}[t]{.45\textwidth}
  \centering
  \caption{\bf QA vs. IP(Urdaneta) in MDRRTs}
    \begin{tabular}{|l|r|r|}
        \hline
        {\bf Teams} &  \bf Breaks(QA) &  \bf Time(Urdaneta) \\
        \hline
        4                       &            6.0 &       0.039347 \\
        \hline
        8                       &           19.6 &       0.062397 \\
        \hline
        12                      &           38.8 &       0.148171 \\
        \hline
        16                      &           66.0 &       0.351599 \\
        \hline
        20                      &          106.8 &       2.837216 \\
        \hline
        24                      &          161.6 &       8.021498 \\
        \hline
        28                      &          224.8 &      26.003237 \\
        \hline
        32                      &          280.4 &      49.271526 \\
        \hline
        36                      &          368.8 &      84.806435 \\
        \hline
        40                      &          453.6 &      75.338228 \\
        \hline
        44                      &          553.6 &      57.133598 \\
        \hline
        48                      &          663.6 &       6.880923 \\
        \hline
    \end{tabular}
    \label{tab:experiment2}
  \end{minipage}
  \hfill
  \begin{minipage}[t]{.45\textwidth}
  \centering
    \caption{\bf QA vs. IP(Urdaneta) in DRRTs}
        \begin{tabular}{|l|r|r|}
            \hline
            {\bf Teams} & \bf Breaks(QA) & \bf Time(Urdaneta) \\
            \hline
            4                       &            5.6 &       0.026426 \\
            \hline
            8                       &           26.8 &       0.055700 \\
            \hline
            12                      &           65.6 &       0.421455 \\
            \hline
            16                      &          116.4 &       5.954419 \\
            \hline
            20                      &          184.4 &      11.851826 \\
            \hline
            24                      &          276.0 &       9.063769 \\
            \hline
            28                      &          409.6 &       1.988316 \\
            \hline
        \end{tabular}
        \label{tab:experiment2_drrt}
    \end{minipage}
\end{tabular}
\end{table}

We now explain the results of the first experiment in the MDRRT (Table \ref{tab:experiment1}). Both IP(Urdaneta) and IP(Trick) took longer to compute as the number of teams increased. Up to 24 teams, IP(Urdaneta) obtained optimal solutions for all problems within 300 \si{\second}. However, in problems with more than 28 teams, the obtained solution was either not optimal or the optimality could not be confirmed. The results of IP(Trick) were inferior to those of IP(Urdaneta). IP(Trick) obtained the optimal solutions within 300 \si{\second} for problems with 16 teams or less. For problems with more than 20 teams, we did not obtain any feasible solutions within 300 \si{\second}. However, QA obtained optimal solutions for problems with 20 teams or less in 0.05 \si{\second}. For problems with 24 teams, the difference between the optimal solution and the solution given by QA was only 5.2. For problems with more than 28 teams, the gap between QA and IP(Urdaneta) gradually became wider. We also conducted the same experiment for DRRTs, and the results are summarized in Table \ref{tab:experiment1_drrt}. Table \ref{tab:experiment1_drrt} shows that the gap between Breaks of QA and IP(Urdaneta) is larger in DRRTs, than in MDRRTs, as the number of teams increased. This is because the break minimization problems in MDRRTs are sparser than in DRRTs, as explained in Section \ref{sec:Analysis:The Benefits of the sparsity of the problem}.

We explain the results of the second experiment as follows (Table \ref{tab:experiment2}). Because the results of IP(Urdaneta) were superior to those of IP(Trick) in the first experiment (Table \ref{tab:experiment1}), we did not conduct the second experiment for IP(Trick).  In the second experiment, we measured the time required for IP(Urdaneta) to reach the objective function value, which QA obtained in 0.05 \si{\second}. \textbf{Time(Urdaneta)} in Table \ref{tab:experiment2} shows that as the number of teams increased, the time taken for IP(Urdaneta) to reach the QA's objective function value also increased, and that the longest time of 84.8 \si{\second} was required for 36 teams. This demonstrates that quantum annealing has better performance than the solver in the break minimization problem in an MDRRT. When the number of teams was 40 or more, the time consumed by IP(Urdaneta) gradually shortened. We consider that this is because the quality of the QA solutions deteriorates when there are a large number of teams. As can be observed from \textbf{Qubits/Nodes} in Table \ref{tab:nodes_edges}, the number of qubits required to represent a variable increases as the number of teams increases. Because the connectivity between qubits is sparse, multiple qubits are required to represent one variable, which may deteriorate the quality of the solutions. By contrast, the experimental results in a DRRT (Table \ref{tab:experiment2_drrt}) show that IP(Urdaneta) takes up to 11.8 \si{\second} to reach the objective function value, which QA obtained in 0.05 s. This result is inferior to that of MDRRTs; therefore, we suggest that the break minimization problem in an MDRRT is more suitable for solving using a quantum annealer, than in a DRRT.

As can be observed from the two experiments, while solving the break minimization problem in an MDRRT, our method that employs the quantum annealer is much faster than integer programming approaches using Gurobi. There are two main reasons for this result. First, as explained in Section \ref{sec:Analysis:The Benefits of the sparsity of the problem} and \ref{sec:Analysis:The Benefits of No Constraints}, the break minimization problem in an MDRRT is sparse and has no constraints; thus, it is easy to solve using a quantum annealer. Second, as Trick \cite{trick2000schedule} highlighted, the break minimization problem is highly symmetric, and it is difficult to solve using an integer programming approach. In addition, our results differ from those of  \cite{ohzeki2019control} in that the quantum annealer is faster as the size of the problems increases. In \cite{ohzeki2019control}, the quantum annealer was faster than Gurobi, but only for small-scale problems.

\section{Conclusion}
\label{sec:Conclusion}

In recent years, with the technical development of quantum annealers, extensive research on solving practical combinatorial optimization problems using quantum annealers has been conducted \cite{neukart2017traffic,nishimura2019item,inoue2021traffic,negre2020detecting,ohzeki2019control,stollenwerk2019quantum}. However, researchers struggle to find practical combinatorial optimization problems, for which quantum annealers outperform other mathematical optimization solvers \cite{ohzeki2019control,o2018nonnegative}. We determined that the break minimization problem in an MDRRT is a problem in which one of the most sophisticated mathematical optimization solvers, such as Gurobi \cite{gurobi} is not suitable, and that quantum annealers are. We formulated the QUBO of the break minimization problem in an MDRRT by referring to existing studies \cite{urdaneta2018alternative}. Further, we used the two superior methods based on integer programming, and our method based on quantum annealing to solve the break minimization problem in an MDRRT, and compared the quality of the solution and computational time. We used Gurobi as an integer programming approach in our experiments. Quantum annealing was able to obtain the exact solution in 0.05 \si{\second} for the problems with 20 teams, which is a practical size. In the case of 36 teams, it took 84.8 \si{\second} for the integer programming method to reach the objective function value, which was obtained by a quantum annealer in 0.05 \si{\second}. The advantage of the method based on quantum annealing is that it is not limited to small-scale problems, which is different from \cite{ohzeki2019control}. Our study is also one of the few to compare quantum annealers with commercial solvers, such as Gurobi.

We provided two primary reasons as to why quantum annealers can successfully solve the break minimization problem in an MDRRT.  First being, that the break minimization problem in an MDRRT has a sparse structure. We demonstrated that the break minimization problem in an MDRRT can be represented as a 4-regular graph. Such a sparse problem is easy to solve using quantum annealers. Second being, that the break minimization problem in an MDRRT is unconstrained.  We highlighted that the unconstrained optimization problem is easy to solve because the number of solutions explored using quantum annealing is equal to the number of feasible solutions. These provide an idea about the types of problems that should be solved using a quantum annealer.

\section*{Acknowledgments}
We would like to thank Prof. Matsui from the Department of Industrial Engineering and Economics, Tokyo Institute of Technology for the valuable advice given to us during the conceptual phase of our research. We would like to thank Editage (www.editage.com) for English language editing. 

Disclaimer: All company or product names mentioned herein are trademarks or registered trademarks of their respective owners.


\begin{thebibliography}{10}

\bibitem{kadowaki1998quantum}
Kadowaki T, Nishimori H.
\newblock Quantum annealing in the transverse Ising model.
\newblock Physical Review E. 1998;58(5):5355.

\bibitem{boothby2020next}
Boothby K, Bunyk P, Raymond J, Roy A.
\newblock Next-generation topology of d-wave quantum processors.
\newblock arXiv preprint arXiv:200300133. 2020;.

\bibitem{dwave2021usermanual}
{D-Wave Systems Inc }. QPU Solver Datasheet; 2021.
\newblock Available from:
  \url{https://docs.dwavesys.com/docs/latest/doc_qpu.html}.

\bibitem{dwave2019noise}
{D-Wave Systems Inc }. Improved coherence leads to gains in quantum annealing
  performance; 2019.
\newblock Available from:
  \url{https://www.dwavesys.com/media/fbpj1x2v/14-1037a-a_improved_coherence_leads_to_gains_qa_performance.pdf}.

\bibitem{neukart2017traffic}
Neukart F, Compostella G, Seidel C, Von~Dollen D, Yarkoni S, Parney B.
\newblock Traffic flow optimization using a quantum annealer.
\newblock Frontiers in ICT. 2017;4:29.

\bibitem{nishimura2019item}
Nishimura N, Tanahashi K, Suganuma K, Miyama MJ, Ohzeki M.
\newblock Item listing optimization for e-commerce websites based on diversity.
\newblock Frontiers in Computer Science. 2019;1:2.

\bibitem{inoue2021traffic}
Inoue D, Okada A, Matsumori T, Aihara K, Yoshida H.
\newblock Traffic signal optimization on a square lattice with quantum
  annealing.
\newblock Scientific reports. 2021;11(1):1--12.

\bibitem{negre2020detecting}
Negre CF, Ushijima-Mwesigwa H, Mniszewski SM.
\newblock Detecting multiple communities using quantum annealing on the D-Wave
  system.
\newblock Plos one. 2020;15(2):e0227538.

\bibitem{ohzeki2019control}
Ohzeki M, Miki A, Miyama MJ, Terabe M.
\newblock Control of automated guided vehicles without collision by quantum
  annealer and digital devices.
\newblock Frontiers in Computer Science. 2019;1:9.

\bibitem{stollenwerk2019quantum}
Stollenwerk T, O’Gorman B, Venturelli D, Mandra S, Rodionova O, Ng H, et~al.
\newblock Quantum annealing applied to de-conflicting optimal trajectories for
  air traffic management.
\newblock IEEE transactions on intelligent transportation systems.
  2019;21(1):285--297.

\bibitem{o2018nonnegative}
O’Malley D, Vesselinov VV, Alexandrov BS, Alexandrov LB.
\newblock Nonnegative/binary matrix factorization with a d-wave quantum
  annealer.
\newblock PloS one. 2018;13(12):e0206653.

\bibitem{gurobi}
{Gurobi Optimization, LLC}. {Gurobi Optimizer Reference Manual}; 2021.
\newblock Available from: \url{https://www.gurobi.com}.

\bibitem{cplex2009v12}
{Cplex, IBM ILOG}.
\newblock V12. 1: User’s Manual for CPLEX.
\newblock International Business Machines Corporation. 2009;46(53):157.

\bibitem{ribeiro2012sports}
Ribeiro CC.
\newblock Sports scheduling: Problems and applications.
\newblock International Transactions in Operational Research.
  2012;19(1-2):201--226.

\bibitem{rasmussen2008round}
Rasmussen RV, Trick MA.
\newblock Round robin scheduling--a survey.
\newblock European Journal of Operational Research. 2008;188(3):617--636.

\bibitem{trick2000schedule}
Trick MA.
\newblock A schedule-then-break approach to sports timetabling.
\newblock In: International Conference on the Practice and Theory of Automated
  Timetabling. Springer; 2000. p. 242--253.

\bibitem{regin2001minimization}
R{\'e}gin JC.
\newblock Minimization of the number of breaks in sports scheduling problems
  using constraint programming.
\newblock DIMACS series in discrete mathematics and theoretical computer
  science. 2001;57:115--130.

\bibitem{urdaneta2018alternative}
Urdaneta HL, Yuan J, Siqueira AS.
\newblock Alternative Integer linear and Quadratic Programming Formulations for
  HA-Assignment Problems.
\newblock Proceeding Series of the Brazilian Society of Computational and
  Applied Mathematics. 2018;6(1).

\bibitem{elf2003minimizing}
Elf M, J{\"u}nger M, Rinaldi G.
\newblock Minimizing breaks by maximizing cuts.
\newblock Operations Research Letters. 2003;31(5):343--349.

\bibitem{miyashiro2006semidefinite}
Miyashiro R, Matsui T.
\newblock Semidefinite programming based approaches to the break minimization
  problem.
\newblock Computers \& Operations Research. 2006;33(7):1975--1982.

\bibitem{suzuka2007home}
Suzuka A, Miyashiro R, Yoshise A, Matsui T.
\newblock The home--away assignment problems and break
  minimization/maximization problems in sports scheduling.
\newblock Pacific Journal of Optimization. 2007;3:113--33.

\bibitem{nemhauser1998scheduling}
Nemhauser GL, Trick MA.
\newblock Scheduling a major college basketball conference.
\newblock Operations research. 1998;46(1):1--8.

\bibitem{schreuder1992combinatorial}
Schreuder JA.
\newblock Combinatorial aspects of construction of competition Dutch
  professional football leagues.
\newblock Discrete Applied Mathematics. 1992;35(3):301--312.

\bibitem{rasmussen2008scheduling}
Rasmussen RV.
\newblock Scheduling a triple round robin tournament for the best Danish soccer
  league.
\newblock European Journal of Operational Research. 2008;185(2):795--810.

\bibitem{ribeiro2006scheduling}
Ribeiro CC, Urrutia S.
\newblock Scheduling the Brazilian soccer tournament with fairness and
  broadcast objectives.
\newblock In: International Conference on the Practice and Theory of Automated
  Timetabling. Springer; 2006. p. 147--157.

\bibitem{kirkpatrick1983optimization}
Kirkpatrick S, Gelatt CD, Vecchi MP.
\newblock Optimization by simulated annealing.
\newblock science. 1983;220(4598):671--680.

\bibitem{de1981scheduling}
De~Werra D.
\newblock Scheduling in sports.
\newblock Studies on graphs and discrete programming. 1981;11:381--395.

\bibitem{dwave2021ocean}
{D-Wave Systems Inc}. dwave-system Documentation Release 1.6.0; 2021.
\newblock Available from:
  \url{https://docs.ocean.dwavesys.com/_/downloads/system/en/stable/pdf/}.

\bibitem{hamerly2019experimental}
Hamerly R, Inagaki T, McMahon PL, Venturelli D, Marandi A, Onodera T, et~al.
\newblock Experimental investigation of performance differences between
  coherent Ising machines and a quantum annealer.
\newblock Science advances. 2019;5(5):eaau0823.

\bibitem{koopmans1957assignment}
Koopmans TC, Beckmann M.
\newblock Assignment problems and the location of economic activities.
\newblock Econometrica: journal of the Econometric Society. 1957; p. 53--76.

\bibitem{dantzig1954solution}
Dantzig G, Fulkerson R, Johnson S.
\newblock Solution of a large-scale traveling-salesman problem.
\newblock Journal of the operations research society of America.
  1954;2(4):393--410.

\bibitem{kuramata2021larger}
Kuramata M, Katsuki R, Nakata K.
\newblock Larger Sparse Quadratic Assignment Problem Optimization Using Quantum
  Annealing and a Bit-Flip Heuristic Algorithm.
\newblock In: 2021 IEEE 8th International Conference on Industrial Engineering
  and Applications (ICIEA). IEEE; 2021. p. 556--565.

\bibitem{kirkman1847on}
Kirkman TP.
\newblock On a problem in combinations.
\newblock Cambridge and Dublin Mathematical Journal. 1847;2:191--204.

\bibitem{cai2014practical}
Cai J, Macready WG, Roy A.
\newblock A practical heuristic for finding graph minors.
\newblock arXiv preprint arXiv:14062741. 2014;.

\end{thebibliography}
\end{document}